\documentclass[12pt]{article}
\usepackage[english]{babel}
\usepackage{amsfonts,amssymb}
\usepackage{amsmath}
\usepackage{latexsym}

\newtheorem{statement}{Statement}[section]
\newtheorem{thm}{Theorem}[section]

\begin{document}

\title{How the permutation of edges of a metric graph affects the number of points moving along the edges}
\author{V.\,L. Chernyshev\footnote{National Research University Higher School of Economics (HSE),
         Myasnitskaya Street, 20, Moscow, 101978, Russia; vchernyshev@hse.ru}, A.\,A. Tolchennikov \footnote{
        A. Ishlinsky Institute for Problems in Mechanics, Russian Academy of Sciences,
        Vernadskogo, 101-1, Moscow, 119526, Russia, tolchennikivaa@gmail.com}}
\date{\today}
\maketitle
\begin{abstract}
We consider a dynamical system on a metric graph, that corresponds to a semiclassical solution of a time-dependent Schr\"{o}dinger equation. We omit all details concerning mathematical physics and work with a purely discrete problem. We find a weak inequality representation for the number of points coming out of the vertex of an arbitrary tree graph.
We apply this construction to an ``H-junction'' graph. We calculate the difference between numbers of moving points corresponding to the permutation of edges. Then we find a symmetrical difference of the number of points moving along the edges of a metric graph.
\end{abstract}

\section{Introduction}
Let us consider a finite metric graph (edges of this graph are regular smooth curves with finite length, e.g., \cite{Kuch_Berk_book}) and the following dynamical system on it. Let one point move along the graph at the initial time. In the interior vertices of the graph, it can be divided as follows: if $k$ points came to the vertex of valence $v$ at the same time, then $v$ points would be released, i.e. one point will correspond to one edge. Reflection occurs in vertices of valence one. Time for passing each individual edge (travel or propagation time) is fixed. It is assumed that there are no turning points on the edges. The problem is to analyze the asymptotic behavior of the number of such points on the graph as time increases. The above discrete formulation is a simplification of the problem that arises in the analysis of semiclassical solutions of the Schr\"{o}dinger equation and in particular in the study of the behavior of Gaussian packets on a metric graph (e.g., article \cite{trsa} and references therein).

Differential equations and analysis on metric graphs continue to attract great interest among mathematicians and physicists. Books \cite{Kuch_Berk_book}, \cite{pokkniga} and references in them can be recommended for interested readers. A number of experts now are engaged in quantum mechanics on graphs, for instance, articles \cite{Robbins}, \cite{Noja}. All necessary definitions related to the study of the statistics of Gaussian packets on a metric graph can be found in article \cite{trsa}.

It was shown in \cite{trsa} that the leading coefficient of the asymptotics for the number of moving points on a finite compact metric graph with increasing time for almost all incommensurate propagation times is determined only by the number of edges, the number of vertices of the graph, the sum and the product of the propagation times  of all edges. The next question arises naturally: what characteristics will determine the following members of an asymptotic expansion? It is impossible to obtain an explicit formula in the general case, but for almost all edge propagation times it can be done, if we construct an asymptotic expansion for the number of lattice points in an expanding simplex of dimension greater than two. Overview of the results associated with this well-known problem can be found in \cite{yau}. In the present article, we turn to the discrete formulation and show how to reduce the problem of finding the number of moving points for a finite tree graph to the number-theoretic problem. To do this, we write an exact formula for the number of points by expressing it in terms of the number of integer nonnegative solutions of weak inequalities, which from a geometric point of view corresponds to simplices. In the calculations we use a function, which can be linked with the Sprague-Grundy function (e.g., \cite{sh_gr}), for some games.

Taking into account the fact that the smaller terms of the asymptotic expansion of the number of lattice points in a simplex are symmetrically included in the asymptotic expansion of the number of moving points, we can consider the difference between the number of points for the two graphs that are identical from topological point of view, but have different propagation times. This means that two graphs have the same number of vertices and set of edge propagation times, but have different order of edges. We apply this approach for a $H$-junction (see \cite{Hj}), i.e. a tree graph ${\Gamma}_H$, with five edges and six vertices, two of which are inner vertices and four have valence one. We consider two variants of composition of ${\Gamma}_H$ from the same set of edges. It is shown that the difference of the number of moving points is of the order $T^3$ and the leading coefficient of the difference between the number of points is explicitly expressed in terms of propagation times of all edges except a ``jumper''. It turns out that the second term of the asymptotic number of points depends on where the initial data is taken, therefore we can consider the symmetric difference of the number of moving points over all possible pairs of internal vertices. It turns out that it is of the order $T^2$ and the leading coefficient is explicitly written out.

Computer experiments have been conducted together with O.\,V. Sobolev, where expressions obtained in section \ref{sect_gl} were calculated directly. The results are in accordance with those obtained analytically.

\section{The transition to a set of weak inequalities}

We introduce the following notation. Let $E(\Gamma) = \{e_i\}_{i=1}^{E}$ be a set of edges of the graph $\Gamma$. Propagation times of the point along the edges $E(\Gamma)$ are, respectively, $\{t_i\}_{i=1}^{E}$. Further we assume that $\{t_i\}_{i=1}^{E}$ are linearly independent over $\mathbb{Q}$.

Suppose that there is a tree $\Gamma$ with the root $A$. Let $A$ be a starting point. Let us recall how dynamics on a graph is constructed. One point moves along the graph at the initial time. In the interior vertices of the graph, it can be divided as follows: if $k$ points came to the vertex of valence $v$ at the same time, then $v$ points will start to move over all the edges (one point on an edge). In vertices of valence one reflection occurs. Edge propagation time for each individual edge is fixed. It is assumed that there are no turning points on the edges.

We want to find the number of points moving along the graph $\Gamma$, at the time $T$. We will find times in which new points have been formed, and then sum the number of new points over such times. Each birth time corresponds to a subtree with root in $A$. The subtree consists of the edges $\{e_{i_1},\ldots,e_{i_L}\}$, which the point has passed before returning to $A$.
A set of times of the form $\{2t_{i_1}n_{i_1}+\ldots + 2t_{i_L}n_{i_L} \}$ corresponds to each subtree (the point has passed $2n_i$ times along the edge $e_i$).
If we want to find the number of times that are less than T, it is necessary to find the number of integer points in the simplex $\{2t_{i_1}n_{i_1}+\ldots + 2t_{i_L}n_{i_L} \le T , n_i > 0 \}$. For instance, for the graph $\Gamma_H$ from section \ref{sect_gl} there will be $10$ subtrees with one root and $10$ subtrees with another root. Thus, the number of points may be represented as a linear combination of numbers of integer points in $20$ expanding simplices. However, if we consider slightly larger simplices and allow coordinates to be zero, then there will be fewer simplices: $11$ instead of $20$. In this section, we will present a formula for the number of points born at the vertex $A$, where all simplices will be given via weak inequalities.

In the following text $n_i \in \mathbb{N} \cup \{0\}$.

Let us define $\#$[system of inequalities on $n_i$] as the number of solutions of the system.

\begin{statement}

The following relation holds:

$$
\left[ n_1t_1 +\ldots n_kt_k \le T, n_i > 0 \ \forall \ i=1,\ldots k\right] =
$$
$$
\sum_{s=0}^k (-1)^{(k-s)} \sum_{\{ i_1,\ldots i_s\} \subset \{1,\ldots k\}} [ n_{i_1}t_{i_1} + \ldots + n_{i_s}t_{i_s} \le T , n_{i_j} \ge 0 \ \forall \ j=1,\ldots s].
$$
\label{st1}
\end{statement}

Proof is elementary, by induction.

Let $N(\Gamma, A, X, T)$ be the number of new points born at the vertex $X \in \Gamma$ to the moment $T$, with the condition that the point have started from the vertex $A \in \Gamma$ at the initial time. Let the vertex $A$ be incident to edges $e_1,\ldots, e_{deg(A)}$, where $deg(A)$ is a valence of $A$. Let $\Gamma_{e_i}$ be maximum subtree without edges $e_j, j \ne i$. Then $\Gamma = \cup_{i=1}^{deg(A)} \Gamma_{e_i}$. By $|\Gamma|$ we define the number of edges of $\Gamma$, i.e. $|E(\Gamma)|$.

\textbf{Definition} $$z(\Gamma, G, A) = (-1)^{|G|} \sum_{\Gamma' \subseteq \Gamma
\text{ including } G \subseteq E(\Gamma')} (-1)^{|\Gamma'|}.$$

\textbf{Definition} $z_i(\Gamma, A) = z(\Gamma_{e_i}, \varnothing, A), 1\le i \le
deg(A).$

Let $G=\{e_{i_1}, \ldots e_{i_k}\}$. Let us define

$$ \#[G] = [2n_{i_1}t_{i_1} + \ldots + 2n_{i_k}t_{i_k} \le T, n_i \ge 0].
$$

\begin{thm}\label{st2}

The following relation holds for the quantity $N(\Gamma, A,A,T)$ :

$$
N(\Gamma, A,A,T) = \sum_{G\subseteq E(\Gamma)} c_G \#[G],
$$

where $c_G$ is defined as follows. We enumerate edges incident to $A$, so that $G$ intersect only $E(\Gamma_{e_1}),\ldots, E(\Gamma_{e_k})$ ($deg(A) = n \ge k$). Then $$ c_G = \prod_{i=1}^k z(\Gamma_{e_i}, G, A) \left[ \sum_{j=0}^{n-k-1} (n-k-j) \sigma_j(z_{k+1},\ldots,z_n) \right], $$ where $\sigma_j$ are elementary symmetric polynomials, $z_i = z_i(\Gamma, A)$.

\end{thm}

\textbf{Proof.}
Let us consider a tree $\Gamma_{e_1}$ such that only one edge $e_1$ is incident to the root. We find $F(\Gamma_{e_1}, A, T)$, being the number of times less than $T$, at which a point returns to the root $A$. A tree corresponds to such time. The tree consists of edges that were engaged in the route. A subtree $\Gamma' \subset \Gamma_{e_1}$ with edges $e_1, e_{i_1}, \ldots e_{i_{k-1}}$ corresponds to a set of times $\{ 2n_1t_1 + 2n_{i_1}t_{i_1} + \ldots + 2n_{i_{k-1}}t_{i_{k-1}} \}$
(here $n_1, n_{i_1},\ldots n_{i_{k-1}} > 0$).

Using the statement \ref{st1}, we obtain:
$$
F(\Gamma_{e_1}, A, T) =\\
$$
$$
\sum_{\Gamma'\subseteq \Gamma_{e_1}} \sum_{s=0}^k (-1)^{k-s}
\sum_{j_1,\ldots j_s \in \{ 1, i_1,\ldots i_{k-1} \}} \#[2n_{j_1}t_{j_1}+\ldots 2n_{j_s}t_{j_s} \le T, n_j \ge 0].
$$

If we interchange the order of summation the right-hand side will have the form of a sum over all subsets of edges $G \subset \{e_1,\ldots e_{E(\Gamma_{e_1})}\}$:
$$
F(\Gamma_{e_1}, A, T) =
\sum_{G\subseteq \{ e_1,\ldots e_{E(\Gamma_{e_1})}\}} \#[G] \ z(\Gamma_{e_1}, G, A).
$$

Suppose now that the vertex $A$ is incident to two edges $e_1, e_2$. $\Gamma = \Gamma_{e_1} \cup
\Gamma_{e_2}$. Then, in the similar way:
$$
F(\Gamma, A, T) =
\sum_{G \subseteq \{e_1,\ldots e_{E(\Gamma)}\}} \#[G]
(-1)^{|G|} \sum_{\Gamma' \subseteq \Gamma \ \text{such that}\ G \subseteq E(\Gamma')} (-1)^{|\Gamma'|}.
$$

We divide the $G = G_1 \cup G_2$, $G_1 = E(\Gamma_{e_1}) \cap G, G_2 = E(\Gamma_{e_2}) \cap
G$, then the last formula will have the form:
$$
F(\Gamma, A, T) = \\
$$
$$
\sum_{G_1\subseteq E(\Gamma_{e_1}), G_2\subseteq E(\Gamma_{e_2})} [G_1 \cup G_2] (-1)^{|G_1|+|G_2|} \\
$$
$$
\sum_{\Gamma' \subseteq \Gamma_{e_1}, \Gamma'' \subseteq \Gamma_{e_2} \text{ �.�. } G_1\subseteq E(\Gamma'),
G_2 \subseteq E(\Gamma'')} (-1)^{|\Gamma'| + |\Gamma''|} =
$$
$$
\sum_{G_1\subseteq E(\Gamma_{e_1}), G_2\subseteq
E(\Gamma_{e_2})} [G_1 \cup G_2] z(\Gamma_{e_1}, G_1, A) z( \Gamma_{e_2}, G_2, A).
$$

Suppose now that $n$ edges $e_1,\ldots,e_n$ leave the vertex $A$ of the graph $\Gamma = \Gamma_{e_1,\ldots e_n}$ :
$\Gamma = \cup_{i=1}^n\Gamma_{e_i}$
Then
$$
N(\Gamma, A, A, T) = \sum_{s=1}^{n-1} (n-s) \sum_{\{i_1,\ldots i_s\} \subseteq \{1,\ldots,n\} }
F(\cup_{j=1}^s \Gamma_{e_{i_j}}, A, T).
$$

This formula is derived from the following considerations. Let it be known that the point that returned at the moment $T$ to the vertex A, had passed $s$ edges of the form $e_1,\ldots e_n$. Then $n-s$ new points are formed at the time $T$. And the inner sum is the number of times of this type.

In this sum, we need to collect the coefficients of functions of the form $\#[G]$.

Suppose, for example, $G\subseteq E(\Gamma_{e_1})$, then the coefficient of $\#[G]$ equals
$$
(n-1) z(\Gamma_{e_1}, G, A) + (n-2) z(\Gamma_{e_1}, G, A)\left( z(\Gamma_{e_2}, \varnothing, A) +
\ldots +
z( \Gamma_{e_n}, \varnothing, A) \right) +
$$
$$
+ (n-3) z(\Gamma_{e_1}, G, A)
\left( z(\Gamma_{e_2}, \varnothing, A) z(\Gamma_{e_3}, \varnothing, A)+\ldots \right) + \ldots =
$$
$$
z(\Gamma_{e_1}, G, A) \left(n-1 + (n-2) \sigma_1(z_2,\ldots,z_n) +
(n-3) \sigma_2(z_2,\ldots,z_n) + \ldots\right) = c_G.
$$

Similarly, we obtain the proof in the general case.

Note that the function $z(\Gamma, \varnothing, A)$ can be calculated recursively. Consider
$$
z^0(\Gamma, A) = \sum_{\Gamma' \subseteq \Gamma, A\in \Gamma'} (-1)^{|\Gamma'|} + 1 = z(\Gamma,\varnothing,A) + 1,
$$
that is, we add a term corresponding to the $\Gamma'$, consisting of a single vertex.

Then
1) If the trees $\Gamma_1$ and $\Gamma_2$ intersect only in a single vertex $A$, then
$$z^0(\Gamma_1 \cup \Gamma_2, A) = z^0(\Gamma_1, A)z^0(\Gamma_2, A),$$

2) If only one edge $(A,B)$ is incident the vertex $A$ of the graph $\Gamma_1$ and $\Gamma_2 = \Gamma_1 \setminus (A,B)$, then
$$z^0(\Gamma_1, A) = 1 - z^0(\Gamma_2, B).$$

Note that the function $z^0_A$ can be interpreted as a denial of the Sprague-Grundy function (e.g., \cite{sh_gr}) for some game, as follows. Consider two players moving down the tree from the root, who make moves in turn. The aim of the game is to make the last move (to get to the vertex of the valence one). Winning strategy: move to the vertices labeled $1$. The first player wins the game, using the right strategy, when only when the root is labeled $0$. From vertices labeled $1$,  only vertices labeled $0$ can be reached. From vertices labeled $0$ there always a possibility to move to $1$, therefore, initiative always belongs to the player in the vertex labeled $0$, and he can make the last move (i.e., get to the vertex of the valence one).

\section{Results for the graph ${\Gamma}_H$}

\label{sect_gl}

Consider a $H-junction$, i.e. a graph ${\Gamma}_H$ consisting of edges $e_1, e_2, e_3,
e_4, e_5$, in which propagation times are $t_1, t_2, t_3, t_4, t_5$ respectively. There are only two vertices with the degree 3 in the graph: $A$, incident to  $e_1,e_2,e_3$, and $B$ incident to $e_3,e_4,e_5$. Here we assume that all $t_i$ are linearly independent over $\mathbb{Q}$.


Integers $n_i$ are non-negative in the following inequalities, unless otherwise stated.

\begin{statement}

The following relation holds:
$$
N(\Gamma_H, A, A, T) = $$
$$
\#[2n_1t_1 + 2n_3t_3 + 2n_4t_4 + 2n_5t_5 \le T] + \#[2n_2t_2 + 2n_3t_3 + 2n_4t_4 + 2n_5t_5 \le T] -
$$
$$
- \#[2n_1t_1 + 2n_4t_4 + 2n_5t_5 \le T] - \#[2n_2t_2 + 2n_4t_4 + 2n_5t_5 \le T] +
$$
$$
+ \#[2n_1t_1 + 2n_2t_2 \le T] + \#[2n_1t_1 \le T] + \#[2n_2t_2 \le T] + const.
$$
\label{st3}
\end{statement}

\textbf{Proof.}

Let us calculate coefficients directly, using the formula of the Statement \ref{st2}, considering that:
$$
z_1(\Gamma_H,A) = z_2(\Gamma_H,A) = -1, z_3(\Gamma_H,A) = 0,
$$
$$
z(\Gamma_{e_1}, \{e_1\}, A) = z(\Gamma_{e_2},\{e_2\},A) = 1,
$$
$$
z(\Gamma_{e_3},\{e_3\},A) = 0,
z(\Gamma_{e_3},\{e_3, e_4\},A) = z(\Gamma_{e_3},\{e_3, e_5\}, A) = 0,
$$
$$
z(\Gamma_{e_3},\{e_4, e_5\},A) = -1, z(\Gamma_{e_3},\{e_3, e_4, e_5\}, A) = 1.
$$

\begin{statement}
The following relation holds:
$$
N(\Gamma,A,B,T) = \#[2n_1t_1 + 2n_2t_2 + 2n_3t_3 + 2n_4t_4 \le T - t_3] +
$$
$$
+ \#[2n_1t_1 + 2n_2t_2 + 2n_3t_3 + 2n_5t_5 \le T - t_3] +
$$
$$
+ \#[2n_1t_1 + 2n_2t_2 + 2n_4t_4 + 2n_5t_5 \le T - t_3] -
$$
$$
- \#[2n_1t_1 + 2n_2t_2 \le T - t_3].
$$
\label{st4}
\end{statement}

\textbf{Proof.} Let us consider the following cases:\\
1) $n_3 > 1$, \\
2) $n_3 = 1$, $n_1=n_2=0$ and $n_4 + n_5 \ne 0$, \\
3) $n_3 = 1$ and $n_1 + n_2 \ne 0.$

Sum of the times from $1$ and $2$ gives $N(\Gamma, B, B,(T - t_3))$ (expression derived from
$N(\Gamma, A,A,T)$ by replacing $e_1$ by $e_4$, $e_2$ by $e_5$, $T$ by $T-t_3$).

Case 3 gives
$$
2\#[2n_1t_1 + 2n_2t_2 \le T- t_3] +
$$
$$
+2\#[2n_1t_1 + 2n_2t_2 + 2n_4t_4 \le T- t_3, n_4>0, n_1 + n_2 >0] +
$$
$$
+ 2\#[2n_1t_1 + 2n_2t_2 + 2n_5t_5 \le T- t_3, n_5>0, n_1 + n_2 >0] +
$$
$$
+ \#[2n_1t_1 + 2n_2t_2 + 2n_4t_4 + 2n_5t_5 \le T- t_3, n_4>0, n_5>0, n_1 + n_2 >0].
$$

If we sum these functions we obtain $N(\Gamma, A,B,T)$.

Let us find an expression for $N(\Gamma_H, A, T)$, i.e. the total number of points on the graph $\Gamma_H$ for the moment $T$ where a point came from the vertex $A$ at the initial time. Obviously, $N(\Gamma_H, A, T) = N(\Gamma_H, A, A, T) + N(\Gamma_H, A, B, T)$.

\textbf{Assumption 1} Numbers $\{t_1,t_2,t_3,t_4,t_5\}$ are linearly independent over $\mathbb{Q}$.
Suppose that for all $1\le m \le 4 $ and all m-element subsets 
$\{s_{i_1}, \ldots, s_{i_m}\} \subset \{t_1,t_2,t_3,t_4,t_5\}$ the following holds:
$$
\# [s_{i_1}n_{i_1} + \ldots + s_{i_m}n_{i_m} \le T]\ =
P_m(s_{i_1},\ldots,s_{i_m})T^4 + R_m(s_{i_1},\ldots,s_{i_m})T^3 + o(T^3).
$$

From the work \cite{Skriganov} it follows that the Assumption $1$ holds for almost all\\
$\{t_1,t_2,t_3,t_4,t_5\}$, but there is no rigorous proof of this statement yet.


\begin{statement}
For times $\{t_1,t_2,t_3,t_4,t_5\}$, satisfying the Assumption 1, the following holds:
$$
a) N(\Gamma_H, A, A, T) =
\frac{1}{4! 2^4} \left( \frac{1}{t_2t_3t_4t_5} + \frac{1}{t_1t_3t_4t_5}\right) T^4 +
$$
$$
\left( R_4(2t_2,2t_3,2t_4,2t_5) + R_4(2t_1,2t_3,2t_4,2t_5) -
\frac{1}{48t_2t_4t_5}- \frac{1}{48t_1t_4t_5} \right)T^3 + o(T^3),
$$

$$
b) N(\Gamma_H, A, B, T) =
\frac{1}{4! 2^4} \left( \frac{1}{t_1t_2t_4t_5} + \frac{1}{t_1t_2t_3t_5} + \frac{1}{t_1t_2t_3t_4}\right) T^4 +
$$
$$
\left( R_4(2t_1,2t_2,2t_4,2t_5) + R_4(2t_1,2t_2,2t_3,2t_5)+ R_4(2t_1,2t_2,2t_3,2t_4) + \right.
$$
$$
\left. -\frac{t_3}{96 t_1t_2t_4t_5} -\frac{1}{96 t_1t_2t_5} -\frac{1}{96 t_1t_2t_4} \right) T^3 + o(T^3),
$$

$$
c) N(\Gamma_H, A, T) =
\frac{1}{4!2^4}\left(\sum_{\{ s_1,s_2,s_3,s_4\} \subset \{t_1,t_2,t_3,t_4,t_5\}} \frac{1}{s_1s_2s_3s_4}\right) +
$$
$$
\left( \sum_{\{ s_1,s_2,s_3,s_4\} \subset \{t_1,t_2,t_3,t_4,t_5\}} R(2s_1,2s_2,2s_3,2s_4) - \right.
$$
$$
\left. -\frac{1}{96}\left( \frac{2}{t_2t_4t_5} + \frac{2}{t_1t_4t_5} +
\frac{t_3}{t_1t_2t_4t_5} + \frac{1}{t_1t_2t_4} + \frac{1}{t_1t_2t_5} \right) \right) T^3 + o(T^3).
$$

\end{statement}

Thus, we can get the leading coefficient for the difference between the number of points on the graphs, obtained from each other by the permutation of edges. For example, if $\Gamma'_H$ is obtained from
$\Gamma_H$ by the permutation of $e_1$ and $e_5$, then the following statement holds.

\begin{statement} For times $\{t_1,t_2,t_3,t_4,t_5\}$,
satisfying the Assumption $1$, it holds that:
$$
N(\Gamma_H, A, T) - N(\Gamma'_H, A, T) =
-\frac{1}{96\ t_2t_4}\left(\frac{1}{t_5} - \frac{1}{t_1}\right) T^3 + o(T^3).
$$
\label{razn}
\end{statement}

 One can note that the difference of the number of moving points is of the order $T^3$ and the leading coefficient of the difference of the number of points is explicitly expressed in terms of propagation times of all edges except a ``jumper'', i.e. $t_3$.

\subsection{Symmetric difference}

\textbf{Assumption 2} Numbers $\{t_1,t_2,t_3,t_4,t_5\}$ are linearly independent over $\mathbb{Q}$.

Suppose that for all $1\le m \le 4 $ and all m-element subsets 
$\{s_{i_1}, \ldots, s_{i_m}\} \subset \{t_1,t_2,t_3,t_4,t_5\}$ the following holds:
$$
\# [s_{i_1}n_{i_1} + \ldots + s_{i_m}n_{i_m} \le T]\ = P_m(s_{i_1},\ldots,s_{i_m})T^4 +
$$
$$
+ R_m(s_{i_1},\ldots,s_{i_m})T^3 + K_m(s_{i_1},\ldots,s_{i_m})T^2 + o(T^2).
$$

Let us denote internal vertices of one variant of the composition of ${\Gamma}_H$ by $A$ and $B$, and of the other variant by $X$ and $Y$. Since the second term of the asymptotic number of points depends on the initial position of the point, we can consider different variants of differences of the number of points depending on which two vertices are fixed. To eliminate the arbitrary choice of pairs, we can take the sum of the differences of the number of points over all possible permutations. It turns out that such symmetric difference is of the order $T^2$, not $T^3$. More precisely, the following proposition holds.

\begin{statement}
If we denote the difference between the number of points issued from the vertex $A$ of the first graph and the number of points issued from the vertex $X$ of the second graph as $d(A, X)$, than for times $\{t_1,t_2,t_3,t_4,t_5\}$, satisfying the Assumption 2, it holds that:
$$
d(A, X) + d(A,Y) + d(B,X) + d(B,Y) = \\
$$
$$
=-\frac{1}{4} \left( \frac{1}{t_1t_2} + \frac{1}{t_4t_5} - \frac{1}{t_2t_4} -
\frac{1}{t_1t_5} \right)T^2+o(T^2).
$$
\label{symraz}
\end{statement}

The proof is analogous to that of the previous statements and comes down to the consideration of a system of weak inequalities.
A computer-based experiment has been carried out, in which expressions st out in the Statements \ref{razn} and \ref{symraz}, were obtained by direct calculation. The graph ${\Gamma}_H$ was taken, with edge propagations times: $t_1=1, t_2=\sqrt{2}, t_3=\sqrt{3}, t_4=\sqrt{5}, t_5=\sqrt{7}$. Points were coming out of one of the two internal vertices at the initial moment.
The results of the experiment are in accordance with the analytical statements. The experiment was carried out in cooperation with O.\,V. Sobolev.

\section{Acknowledgements}

The authors are grateful to A.\,I. Esterov, M.\,M. Skriganov, N.\,G. Moschevitin, O.\,N. German, A.\,V. Gorbunov, O.\,V. Sobolev and A.\,I. Shafarevich for useful discussions and attention to their work.

\label{end}

\end{document}